\documentclass[12pt]{article}
\usepackage{times}
\usepackage{amsmath}
\usepackage{amssymb}
\usepackage{amsthm}
\usepackage{booktabs}
\usepackage{graphicx}
\usepackage{subfigure}
\usepackage[super, numbers,sort&compress]{natbib}
\usepackage{setspace}
\usepackage[hmargin=1.1in,vmargin=1.1in]{geometry}
\usepackage{xcolor}
\usepackage{amsfonts}
\usepackage{lipsum}
\usepackage{authblk}
\usepackage{url}
\usepackage{indentfirst}
\usepackage{makecell}

\usepackage{multirow}

\usepackage[algoruled,boxed,lined]{algorithm2e}
\usepackage{algpseudocode}
\usepackage{bm}
\usepackage{bbm}
\usepackage{mathtools}
\usepackage{setspace}
\usepackage{enumerate}

\usepackage{tikz}

\usetikzlibrary{shapes.geometric,arrows}
\begin{document}
\begin{center}
    \large \bf Re-evaluating the impact of reduced malaria prevalence on birthweight in sub-Saharan Africa: A pair-of-pairs study via two-stage bipartite and non-bipartite matching
\end{center}

\begin{center}
    {\large $\text{Pengyun Wang}^{1}$, $\text{Ping Huang}^{2}$, $\text{Yifan Jin}^{3}$, $\text{Yanxin Shen}^{4}$,\\ $\text{Omar El Shahawy}^{5}$, $\text{Dae Woong Ham}^{3}$, $\text{Wendy P. O'Meara}^{6}$, and $\text{Siyu Heng}\textsuperscript{5, \dag}$}
  \end{center}
  \begin{center}
     \large $^{1}\text{\textit{University of Oxford}}$, $^{2}\text{\textit{Peking University}}$, $^{3}\text{\textit{University of Michigan}}$, $^{4}\text{\textit{Nankai University}}$, $^{5}\text{\textit{New York University}}$, and $^{6}\text{\textit{Duke University}}$
  \end{center}


\let\thefootnote\relax\footnotetext{$^{\dagger}$ Corresponding Author: Siyu Heng, Department of Biostatistics, School of Global Public Health, New York University, New York, NY 10003, United States (email: siyuheng@nyu.edu). }

\let\thefootnote\relax\footnotetext{$^{*}$ This manuscript has not been peer-reviewed and is pending submission. }

\begin{abstract}

According to the World Health Organization (WHO), in 2021, about 32\% of pregnant women in sub-Saharan Africa were infected with malaria during pregnancy. Malaria infection during pregnancy can cause various adverse birth outcomes such as low birthweight. Over the past two decades, while some sub-Saharan African areas have experienced a large reduction in malaria prevalence due to improved malaria control and treatments, others have observed little change. Individual-level interventional studies have shown that preventing malaria infection during pregnancy can improve birth outcomes such as birthweight; however, it is still unclear whether natural reductions in malaria prevalence may help improve community-level birth outcomes. We conduct an observational study using 203{,}141 children’s records in 18 sub-Saharan African countries from 2000 to 2018. Using heterogeneity of changes in malaria prevalence, we propose and apply a novel pair-of-pairs design via two-stage bipartite and non-bipartite matching to conduct a difference-in-differences study with a continuous measure of malaria prevalence, namely the \textit{Plasmodium falciparum} parasite rate among children aged 2 to 10 ($\text{\textit{Pf}PR}_{2-10}$). The proposed novel statistical methodology allows us to apply difference-in-differences without dichotomizing $\text{\textit{Pf}PR}_{2-10}$, which can substantially increase the effective sample size, improve covariate balance, and facilitate the dose-response relationship during effect estimation. Our outcome analysis finds that among the pairs of clusters we study, the largest reduction in $\text{\textit{Pf}PR}_{2-10}$ over early and late years is estimated to increase the average birthweight by 98.899 grams (95\% CI: $[39.002, 158.796]$), which is associated with reduced risks of several adverse birth or life-course outcomes. Our findings support future studies and investments in mitigating the global malaria burden. The proposed novel statistical methodology can be replicated in many other disease areas. 

\end{abstract}
\noindent%

\section{Introduction}\label{sec:intro}

Malaria is a parasitic disease that has posed a significant global health challenge~\cite{assefa2023universal, poespoprodjo2023malaria}. Between 2000 and 2022, approximately 2 billion malaria cases occur worldwide, leading to about 12 million deaths~\cite{venkatesan20242023}. In particular, 82\% of malaria cases and 95\% of the resulting deaths occurred in sub-Saharan Africa~\cite{gilmartin2021seasonal, venkatesan20242023}. Pregnant women are especially vulnerable to malaria infection due to both generalized immune suppression and sequestration of parasitized cells in the placenta~\cite{desai2007epidemiology, rogerson2007malaria}. In 2021, about 32\% of pregnant women in sub-Saharan Africa were infected with malaria during pregnancy \cite{WHO2022, monroe2022reflections}. Malaria infection during pregnancy can cause serious adverse birth outcomes~\cite{heng2021relationship}, where malaria parasites sequestered in the placenta interfere with blood flow to the fetus and ultimately starve the fetus of nutrients~\cite{desai2007epidemiology, de2013impact, gilmartin2021seasonal}. This cascade increases the risk of adverse birth outcomes such as low birthweight, preterm delivery, neonatal death, and stillbirth~\cite{heng2021relationship, desai2007epidemiology, de2013impact}.

Malaria control has been successful in sub-Saharan Africa, where, from 2000 to 2022, malaria cases are estimated to have declined by 41\%~\cite{venkatesan20242023}. Although previous individual interventional studies \cite{kayentao2013intermittent} have found that preventing malaria infection during pregnancy can reduce the risk of adverse birth outcomes such as low birthweight \textit{at the individual level}, it is still unclear if such a significant reduction in malaria transmission rates would help reduce instances of adverse birth outcomes \textit{at the community level}. This is because reducing malaria prevalence at the community level will also reduce the community's antimalarial immunity, including maternal antimalarial immunity~\cite{rogerson2007malaria}; maternal antimalarial immunity is important in mitigating the adverse impacts of malaria infection during pregnancy~\cite{mayor2015changing}. This is also evidenced by the reduced impacts of malaria infection in second, third, and subsequent pregnancies among infected mothers~\cite{accrombessi2019effects, cutts2020pregnancyspecific}. Therefore, it is unclear whether the overall reduction in exposure to malaria infection during pregnancy (due to reduced malaria prevalence) outweighs the increased individual risk (due to reduced maternal antimalarial immunity caused by reduced malaria prevalence) of adverse birth outcomes once infected; see Appendix D for more details. 

In this study, we concentrate on birthweights, given their role as a key indicator of neonatal health and their potential to serve as proxies for various other infant health metrics, particularly in assessing the community-level impacts of reduced malaria prevalence~\cite{mcclure2013systematic, rijken2014quantifying}. Heng et al. (2021)\cite{heng2021relationship} conducted an observational study on this topic using a novel study design called pair-of-pairs, which combines the difference-in-differences and matching methods to compare the low birthweight rate trajectories among sub-Saharan Africa regions with substantially reduced malaria prevalence (i.e., the exposed group) and those with consistently high malaria prevalence (i.e., the control group) \citealp{bhatt2015effect, bejon2010stable}. However, the statistical analysis in Heng et al. (2021)\cite{heng2021relationship} failed to detect strong evidence of the community-level effect of reduced malaria prevalence on increased birthweights in sub-Saharan Africa. We hypothesize that this is due to a significant limitation of the study design adopted in Heng et al. (2021)\cite{heng2021relationship} -- the originally continuous measure of malaria prevalence needs to be empirically and somewhat arbitrarily dichotomized into high (i.e., the control) or low malaria prevalence (i.e., the intervention) to facilitate a difference-in-difference study, which can render three issues: reduced effective sample size due to dichotomization before statistical matching, suboptimal covariate imbalance after statistical matching, and most importantly, bias due to ignoring dose-response relationship during outcome analysis. 

To re-evaluate the community-level impacts of reduced malaria prevalence on infants' birthweights in sub-Saharan Africa, we conduct an observational study among 203{,}141 children's records in 18 sub-Saharan African countries. Our study proposes and uses a novel statistical approach -- a pair-of-pairs design via two-stage bipartite and non-bipartite matching -- to conduct a difference-in-differences study with a continuous measure of malaria prevalence. Compared with the previous pair-of-pairs design proposed in Heng et al. (2021)\cite{heng2021relationship}, our proposed approach does not require dichotomizatizing the continuous malaria prevalence rate before statistical matching and, therefore, can retain the exact malaria prevalence rates in statistical analysis. This can lead to over ten times larger effective sample size using similar data resources (203{,}141 effective individual records in our study versus 18{,}499 effective individual records in Heng et al. (2021)), substantially improved covariate balance, and, most importantly, allow effect size to scale with changes in malaria prevalence (i.e., dose-response relationship).

\section{Methods}\label{sec: methods}
\subsection{Step 0: Data Sources and Data Selection Procedure} \label{subsec: Data}
Our study combines the following three rich data sources: 
\begin{itemize}
    \item Rasters of annual malaria prevalence from the Malaria Atlas Project (MAP): These rasters data formulated by the MAP~\cite{MAP} depict the annual spatial distribution of the \textit{Plasmodium falciparum} (one of the most common malaria parasites) parasite rate among children aged 2 to 10 (denoted as $\text{\textit{Pf}PR}_{2-10}$) in sub-Saharan African, spanning from 2000 to 2018. The $\text{\textit{Pf}PR}_{2-10}$ is one of the most commonly used measures for overall malaria transmission intensity \cite{hay2006malaria} because children aged 2 to 10 years are typically exposed to malaria transmission environments without sufficient immunity~\cite{snow2005global}; their infection rate can effectively reflect the malaria endemicity in a given area~\cite{MAP, amoah2021population}. Each raster pixel in the MAP data encapsulates the estimated annual $\text{\textit{Pf}PR}_{2-10}$, varying between 0 and 1 (from the lowest to the highest rate), with a spatial granularity of 5km by 5km. 
    
    \item Demographic and Health Surveys (DHS): The DHS~\cite{corsi2012demographic, ipumsdhs2019} encompasses a series of nationally representative household surveys predominantly executed in low to middle-income nations. These surveys capture a plethora of health, sociodemographic, and socioeconomic metrics~\cite{corsi2012demographic, dhs}. We employ the Integrated Public Use Microdata Series's recoding of the DHS datasets (IPUMS-DHS), which ensures consistency in DHS variables across diverse survey iterations and timelines~\cite{ipumsdhs2019}.
    
    \item Cluster Global Positioning System (GPS): The cluster GPS datasets encapsulate geographical attributes (longitude, latitude, and urban/rural indicators) for each cluster in the DHS data. The spatial deviation between each record in the GPS data and the actual location of the corresponding DHS cluster does not exceed 10 kilometers -- with over 99\% of them having an error margin below 5 kilometers~\cite{dhs}.
\end{itemize}

Our study encompasses all sub-Saharan African countries that meet the following two criteria: (i) The country is covered by the estimated annual $\text{\textit{Pf}PR}_{2-10}$ rasters generated by the MAP between 2000 and 2018, and (ii) the country has at least one standard DHS dataset in IPUMS-DHS from 2000-2010 (``early-years") and at least one from 2011-2018 (``late-years"), and both include cluster GPS coordinates. If multiple DHS early (late) years satisfy these two criteria for a country, we select the earliest early year and the latest late year to maximize the time gap, which offers sufficient time to observe changes in community-level malaria prevalence and detect resulting changes in birth outcomes. Additionally, countries with at least one standard DHS with available cluster GPS data in the late years (2011-2018) but lacking such data in the early years (2000-2010) are still included if they had corresponding DHS data in 1999. In this case, we assign the 2000 MAP data to the 1999 DHS data for such countries, allowing the inclusion of two additional countries, Cote d'Ivoire and Tanzania. The 18 sub-Saharan African countries that satisfy these criteria are listed in Table~\ref{tab: years}. 

\subsection{Step 1 (Stage-One Matching): Using Optimal Bipartite Matching to Pair Early-Year and Late-Year Survey Clusters}\label{subsec: step 1}

We estimate the causal effect of reduced malaria prevalence on birthweight in sub-Saharan Africa using a difference-in-differences study with continuous exposure (i.e., $\text{\textit{Pf}PR}_{2-10}$). In a traditional difference-in-differences study with binary exposure, the exposure effect is calculated as the difference in mean change in outcomes (between the early and late year) among exposed units and that among control units ~\cite{athey2006identification, imbens2009recent, angrist2009mostly}. Similarly, in a difference-in-differences study with continuous exposure, exposure effects can be calculated as the ratio of (i) the difference in mean change in outcomes (between the early and late year) among units (e.g., DHS clusters) with bigger changes in exposure dose (hence force called the ``bigger-exposure-change units") and that among the units with smaller changes (hence force called the ``smaller-exposure-change units") and (ii) the difference in mean change in exposures among the bigger-exposure-change units and that among the smaller-exposure-change units \citealp{callaway2024difference}. 

Using a difference-in-differences approach can remove the following two sources of bias ~\cite{bertrand2004much, heng2021relationship}: First, the difference-in-differences approach can remove bias due to baseline differences in average birthweights between the bigger-exposure-change units and the smaller-exposure-change units at the beginning of the study period~\cite{ham2024benefits}. In contrast, if the average birthweights among the bigger-exposure-change and those among the smaller-exposure-change units were compared only in the early or late year, these baseline differences could be mistaken for exposure effects~\cite{wing2018designing, ham2024benefits}. Second, the difference-in-differences approach can remove bias from unobserved confounders that are not time-varying. Specifically, under the parallel trend assumption (possibly conditional on observed covariates being adjusted for), any confounding in unobserved covariates, e.g., different altitudes in bigger-exposure-change and smaller-exposure-change units (which could also impact birthweight~\cite{unger1988altitude}), does not contaminate or bias our final estimated effect because they ``cancel out" in a difference-in-differences study ~\cite{card2000minimum, bertrand2004much, wing2018designing, ham2024benefits}.

However, within each survey country, the DHS conducts surveys at different survey clusters (i.e., survey locations) over different survey years~\cite{ipumsdhs2019} (i.e., time and location heterogeneity). This fact makes the traditional difference-in-differences study inapplicable because the traditional difference-in-differences requires the same cluster to be measured twice over different survey years (one in the early years and one in the late years). Following the strategy proposed in Heng et al. (2021) \cite{heng2021relationship}, to construct pairs of early-year and late-year DHS clusters that are geographically close such that each pair of clusters can mimic a single cluster surveyed twice in two different years, we use optimal bipartite matching~\cite{rosenbaum1989optimal, hansen2006optimal} to pair clusters from the early years with those from late years based on geographical closeness (measured by latitude and longitude information in the cluster GPS datasets). Specifically, we minimize the total rank-based Mahalanobis distance~\cite{hallin2002optimal} based on the latitude and longitude of the DHS clusters to optimally form paired early-year and late-year clusters based on geographic proximity~\cite{rosenbaum1989optimal, rosenbaum2010design, hansen2006optimal}. Within each country, the number of clusters to pair is set to be the smaller one among the number of clusters in the early year and that in the late year. To ensure geographical closeness for each matched pair of clusters, we eliminate the 505 matched pairs of early-year and late-year clusters between which the spherical distance exceeds 100 km. We also exclude 207 pairs (out of 5{,}368) clusters where both early-year and late-year $\text{\textit{Pf}PR}_{2-10}$ values are zero. This is because these high-altitude areas or economically developed urban regions have historically lacked conditions conducive to malaria transmission~\cite{baragatti2009social, bakken2021impact}. Consequently, variations in birthweight in these regions are unrelated to malaria prevalence. 

After removing those pairs of clusters with sharp transitions in $\text{\textit{Pf}PR}_{2-10}$ due to extreme altitudes or economic conditions (leaving us 5{,}161 pairs of clusters), we assume that the spatial distributions of $\text{\textit{Pf}PR}_{2-10}$ and average birthweight are continuous functions of longitude and latitude (this assumption will be partially assessed in Section~\ref{subsec: results of stage-one matching}). Under this assumption, if the two paired early-year and late-year clusters are geographically sufficiently close, the associated $\text{\textit{Pf}PR}_{2-10}$ and birthweight are also close in both the early and late years. Specifically, let cluster $s$ denote the cluster we observe in late years (i.e., $Z_s^{\text{late}}$ and $Y_s^{\text{late}}$ are observed, where $Z_{s}^{\text{late}}$ and $Y_s^{\text{late}}$ denote the $\text{\textit{Pf}PR}_{2-10}$ and average birthweight of cluster $s$ in late years, respectively). For any cluster $s \in \mathcal{S}_{\text{late}}$, where $\mathcal{S}_{\text{late}}$ is the set of all observed clusters in late years, the $Z_s^{\text{early}}$ and $Y_s^{\text{early}}$ (i.e., the $\text{\textit{Pf}PR}_{2-10}$ and average birthweight of cluster $s$ in early years) are missing since we only observe cluster $s$ in late years. In a typical difference-in-differences study, we need to observe all four values of $Z_s^{\text{early}}, Y_s^{\text{early}}, Z_s^{\text{late}}, Y_s^{\text{late}}$. Consequently, to impute the unobserved $Z_s^{\text{early}}$ and $Y_s^{\text{early}}$, we find another cluster $s' \in \mathcal{S}_{\text{early}}$ such that $s$ and $s^{\prime}$ are geographically close (i.e., have similar longitudes and latitudes, as well as similar altitudes). The idea is that after we pair early-year and late-year clusters that are geographically sufficiently close (which will be assessed in Section~\ref{subsec: results of stage-one matching}), we can just treat the two paired clusters as the ``same'' cluster measured twice across time. Formally, we assume that for any $s \in \mathcal{S}_{\text{late}}$ and $s' \in \mathcal{S}_{\text{early}}$ such that $s$ and $s^{\prime}$ are geographically sufficiently close, we have $Z_{s}^{\text{early}} \approx Z_{s^{\prime} }^{\text{early}}, Z_{s}^{\text{late}} \approx Z_{s^{\prime} }^{\text{late}}, Y_{s}^{\text{early}} \approx Y_{s^{\prime} }^{\text{early}}, Y_{s}^{\text{late}} \approx Y_{s^{\prime} }^{\text{late}}$.

\subsection{Step 2 (Stage-Two Matching): Using Optimal Non-Bipartite Matching to Form Pairs of Pairs with Similar Covariates but Unequal Changes in Malaria Prevalence}\label{subsec: nbp matching}

Instead of simply conducting a difference-in-differences analysis after stage-one matching based on the parallel trend assumption (which implies that clusters intervened with the same level of change in exposure dose across time have the same expected level of change in outcome across time ~\cite{card2000minimum, bertrand2004much, donald2007inference, angrist2009mostly, callaway2024difference}), we perform additional stage-two matching to pair bigger-exposure-change and smaller-exposure-change clusters with the same or similar observed covariates in both the early and late years (i.e., the same or similar covariates trajectories). Due to stage-two matching, instead of assuming that the parallel trend assumption holds regardless of covariate values, we only need to assume that the parallel trend assumption holds among the clusters with the same covariates trajectories (i.e., the conditional parallel trend assumption \citealp{dehejia1999causal, basu2020constructing, kahn2020promise, heng2021relationship, ham2024benefits}), of which the explicit form will be stated in (\ref{eq:conditional_PT}).

Specifically, in stage-two matching, we use an optimal non-bipartite matching algorithm ~\cite{lu2001matching, baiocchi2010building, lu2011optimal} to pair pairs of early-year and late-year clusters (formed in stage-one matching) with similar covariates (in both early and late years) but unequal changes in $\text{\textit{Pf}PR}_{2-10}$ across early and late years. Unlike the optimal bipartite matching algorithm adopted in stage-two matching of Heng et al. (2021)\cite{heng2021relationship}, optimal non-bipartite matching does not require dichotomizing $\text{\textit{Pf}PR}_{2-10}$ into high level (control) and low level (intervention) to facilitate matching, which allows a cluster to be paired with any other cluster with optimally similar covariates but unequal change in $\text{\textit{Pf}PR}_{2-10}$. This flexibility offers three key advantages. First, using non-bipartite matching allows the effect size on change in outcomes to scale with change in $\text{\textit{Pf}PR}_{2-10}$~\cite{lu2011optimal}. In contrast, bipartite matching adopted by Heng et al. (2021) \cite{heng2021relationship} needs to assume birthweights do not change with $\text{\textit{Pf}PR}_{2-10}$ once the $\text{\textit{Pf}PR}_{2-10}$ surpasses some threshold (e.g., 0.4) or falls below some threshold (e.g., 0.2), which has been shown to be impractical by previous studies~\cite{eisele2012malaria, ngai2020malaria, heng2021relationship}. Second, optimal non-bipartite matching can retain as many observations as possible (99.4\% in our case), whereas bipartite matching after dichotomization requires researchers to discard observations falling between the thresholds. For example, the analysis in Heng et al. (2021) \cite{heng2021relationship} needs to discard all the clusters with $\text{\textit{Pf}PR}_{2-10}$ between the thresholds 0.2 and 0.4, which results in over 86\% loss of clusters. Third, using a non-bipartite matching procedure can significantly improve covariate balance compared with using bipartite matching (as will be shown in Section~\ref{subsec: stage-two matching}), which will greatly mitigate effect estimation bias due to covariate imbalance~\cite{ham2024benefits}.

\begin{figure}
\caption{Illustration of our two-stage matching procedure based on bipartite and non-bipartite matching. Bigger-exposure-change pairs donate the pairs of early-year and late-year clusters that have experienced a sharper reduction in $\text{\textit{Pf}PR}_{2-10}$ between early and late years. Smaller-exposure-change pairs donate the pairs of early-year and late-year clusters that have seen a less evident change in $\text{\textit{Pf}PR}_{2-10}$ between early and late years.}
\begin{center}
\includegraphics[width=0.9\linewidth]{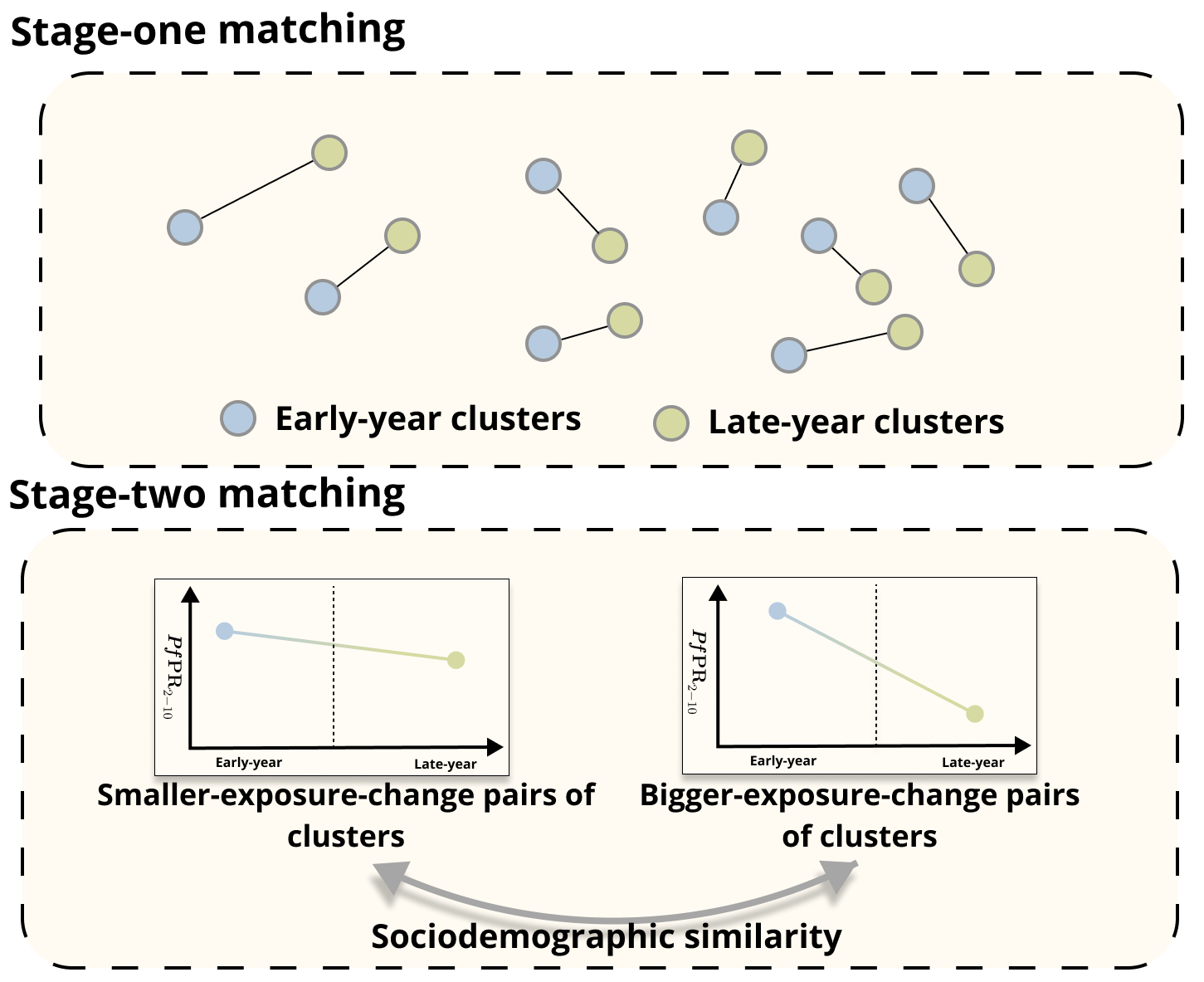}
\end{center}
\label{fig: Quadruples (pairs of pairs) of matched clusters}
\end{figure}

The 12 cluster-level sociodemographic and socioeconomic and socioeconomics covariates (recorded in both early and late years) considered in our non-bipartite matching procedure are listed in Table~\ref{tab:variables_step2}. They are potentially correlated with both the risk of malaria \cite{baragatti2009social, krefis2010principal, bhatt2015effect, cowman2016malaria, roberts2016risk, sulyok2017dsm265, heng2021relationship} and birth outcomes \cite{de2018consumption, gemperli2004spatial, smith2011socioeconomic, jardine2021adverse, padhi2015risk}. These cluster-level covariates are calculated as the within-cluster average values from individual records \cite{kennedy2011adolescent, larsen2017individual}. The missingness rates of these 12 covariates among the individual records are less than 0.1\%. An illustration of our two-stage matching procedure is also presented in Figure~\ref{fig: Quadruples (pairs of pairs) of matched clusters}.

After two-stage matching, suppose that $I$ pairs of pairs (quadruples) of clusters are formed (in our dataset, $I=2{,}565$). For the $j$-th pair of early-year and late-year clusters (formed in stage-one matching) in the $i$-th pair of pairs of early-year and late-year clusters (formed in stage-two matching) ($i=1,\dots, I$ and $j=1,2$), we let $Z_{ij}^{\text{early}}$ (or $Z_{ij}^{\text{late}}$) denote the $\text{\textit{Pf}PR}_{2-10}$ in the early-year (or in the late year), $\mathbf{X}_{ij}^{\text{early}}$ (or $\mathbf{X}_{ij}^{\text{late}}$) denote the 12 covariates considered in stage-two matching in the early-year (or in the late year), and $Y_{ij}^{\text{early}}$ (or $Y_{ij}^{\text{late}}$) denote the cluster-level average birthweight in grams in the early-year (or in the late year). Formally, the conditional parallel trend assumption in our setting can be stated as:
\begin{equation}\label{eq:conditional_PT}
\begin{aligned}
       &\quad \ E[Y_{i1}^{\text{late}}(Z_{i2}^{\text{late}})-Y_{i1}^{\text{early}}(Z_{i2}^{\text{early}}) \mid \mathbf{X}_{i1}^{\text{early}}= \mathbf{X}_{i2}^{\text{early}}, \mathbf{X}_{i1}^{\text{late}}= \mathbf{X}_{i2}^{\text{late}}]\\
      &=E[Y_{i2}^{\text{late}}(Z_{i2}^{\text{late}})-Y_{i2}^{\text{early}}(Z_{i2}^{\text{early}}) \mid \mathbf{X}_{i1}^{\text{early}}= \mathbf{X}_{i2}^{\text{early}}, \mathbf{X}_{i1}^{\text{late}}= \mathbf{X}_{i2}^{\text{late}}],
\end{aligned}
\end{equation}
and symmetrically,
\begin{equation*}
\begin{aligned}
       &\quad \ E[Y_{i1}^{\text{late}}(Z_{i1}^{\text{late}})-Y_{i1}^{\text{early}}(Z_{i1}^{\text{early}}) \mid \mathbf{X}_{i1}^{\text{early}}= \mathbf{X}_{i2}^{\text{early}}, \mathbf{X}_{i1}^{\text{late}}= \mathbf{X}_{i2}^{\text{late}}]\\
      &=E[Y_{i2}^{\text{late}}(Z_{i1}^{\text{late}})-Y_{i2}^{\text{early}}(Z_{i1}^{\text{early}}) \mid \mathbf{X}_{i1}^{\text{early}}= \mathbf{X}_{i2}^{\text{early}}, \mathbf{X}_{i1}^{\text{late}}= \mathbf{X}_{i2}^{\text{late}}],
\end{aligned}
\end{equation*}
in which $Y_{i1}^{\text{late}}(Z_{i2}^{\text{late}})$ and $Y_{i1}^{\text{early}}(Z_{i2}^{\text{early}})$ denote the counterfactual outcome that pair of clusters $i1$ would exhibit if the $\text{\textit{Pf}PR}_{2-10}$ was set to $Z_{i2}^{\text{late}}$ in the late year and that if the $\text{\textit{Pf}PR}_{2-10}$ was set to $Z_{i2}^{\text{early}}$ in the early year (the $Y_{i2}^{\text{late}}(Z_{i1}^{\text{late}})$ and $Y_{i2}^{\text{early}}(Z_{i1}^{\text{early}})$ can be similarly defined), and $Y_{i2}^{\text{late}}(Z_{i2}^{\text{late}})$ and $Y_{i2}^{\text{early}}(Z_{i2}^{\text{early}})$ denote the observed outcome of pair of clusters $i2$ under the observed $\text{\textit{Pf}PR}_{2-10} = Z_{i2}^{\text{late}}$ in the late year and that under the observed $\text{\textit{Pf}PR}_{2-10} = Z_{i2}^{\text{early}}$ in the early year (the $Y_{i1}^{\text{late}}(Z_{i1}^{\text{late}})$ and $Y_{i1}^{\text{early}}(Z_{i1}^{\text{early}})$ can be similarly defined). Equation~\eqref{eq:conditional_PT} and its symmetric form (i.e., the conditional parallel trend assumption) are more likely to hold than the unconditional version (without conditioning on $\mathbf{X}_{i1}^{\text{early}}= \mathbf{X}_{i2}^{\text{early}}$ and $\mathbf{X}_{i1}^{\text{late}}= \mathbf{X}_{i2}^{\text{late}}$). In other words, without matching/controlling for the covariates, the unconditional parallel trends assumption may not hold on the full data set. For example, existing studies have shown that the probability of malaria infection is higher in economically disadvantaged households compared to wealthy households~\cite{krefis2010principal, tusting2013socioeconomic}. Meanwhile, households with different socioeconomic statuses are expected to exhibit significant differences in children's birthweight \cite{ngandu2020association}. Therefore, through matching on various sociodemographic and socioeconomic covariates such as household wealth index, our stage-two matching ensures that the conditional parallel trend assumption holds as much as possible~\cite{heng2021relationship, ham2024benefits}.

\subsection{Step 3: Multiple Imputation After Matching to Address Missingness in Birthweight Records}\label{sec: imputation}

From the initial set of 216{,}779 individual (children) records derived from the 2{,}565 matched pairs of pairs (quadruples) of DHS clusters formed by our two-stage matching procedure, we eliminate those of multiple births (e.g., twins and triplets). This is because previous research has found that the average birthweight among multiple births is typically 600-900 grams lower than that of singletons~\cite{blondel2002impact, elster2000less}. This refinement leaves us with 208{,}789 (96.3\%) individual records. Our outcome of interest is birthweight in grams, of which 50.16\% (104,730) records are missing. To handle this, we perform multiple imputation with 20 replications~\cite{rubin1996multiple, heitjan1996distinguishing, chen2014note, heng2021relationship}. A useful variable for imputing the missing birthweights is the mother's subject report of the child's birth size, of which the efficacy in predicting missing birthweight data has been verified by previous studies ~\cite{blanc2005monitoring}. To ensure imputation quality, we omit 13,638 (6.5\%) records lacking the mother's subjective report on the child's birth size, which leaves us with 203{,}141 records, among which 48.86\% (99{,}253 records) have missing birthweights. We use a Bayesian linear regression with the default weakly informative priors in the \texttt{brm} function (part of the \texttt{brms} package in \texttt{R}) as our imputation model, with the predictors being the covariates (listed in Table S1 in the supplementary materials) that may be associated with both missingness and the actual birthweight, according to previous studies \cite{tyrrell2016height, leroy2015using, heng2021relationship}.

\subsection{Step 4: Estimating the Impact of Reduced Malaria Prevalence on Infants' Birthweights}\label{subsec: methods of missing data imputation}

After two-stage matching and imputation, we are ready to conduct the outcome analysis, which uses a working model to further adjust for residual bias due to imperfect matching. Let $Z_{ij}^{\text{diff}}=Z_{ij}^{\text{late}}-Z_{ij}^{\text{early}}$ denote the difference in $\text{\textit{Pf}PR}_{2-10}$ between the late and early-years (e.g., $Z_{ij}^{\text{diff}}=-0.1$ means the annual $\text{\textit{Pf}PR}_{2-10}$ is reduced by $0.1=10\%$ from the early to late year) and $Y_{ij}^{\text{diff}}=Y_{ij}^{\text{late}}-Y_{ij}^{\text{early}}$ denote the difference in cluster-level average birthweight in grams between the late and early years. Then, we consider the following post-matching working model for outcome analysis:
\begin{equation}\label{eqn: outcome model}
    Y_{ij}^{\text{diff}}= \beta_{0} +\beta_{1}Z_{ij}^{\text{diff}} + \gamma_{\text{early}}^{T}\mathbf{X}_{ij}^{\text{early}} + \gamma_{\text{late}}^{T}\mathbf{X}_{ij}^{\text{late}} + \sum_{i^{\prime}=1}^{I}c_{i^{\prime}} \mathbbm{1}\{i=i^{\prime}\} + \epsilon_{ij},
\end{equation}
where $\mathbbm{1}\{\cdot\}$ is the matched set indicator (i.e., the dummy variable for indicating the $i$-th pair of pairs) and $\epsilon_{ij}$ is the random error term. Note that the validity of a difference-in-differences estimator can be justified by parallel trend assumption alone and does not necessarily require model (\ref{eqn: outcome model}) to hold. Therefore, model (\ref{eqn: outcome model}) mainly serves as a working model for adjusting for residual bias due to imperfect matching on observed covariates after stage-two matching.

\begin{figure} 
\caption{Workflow diagram of the study.}
    \centering
    \includegraphics[width=0.95\linewidth]{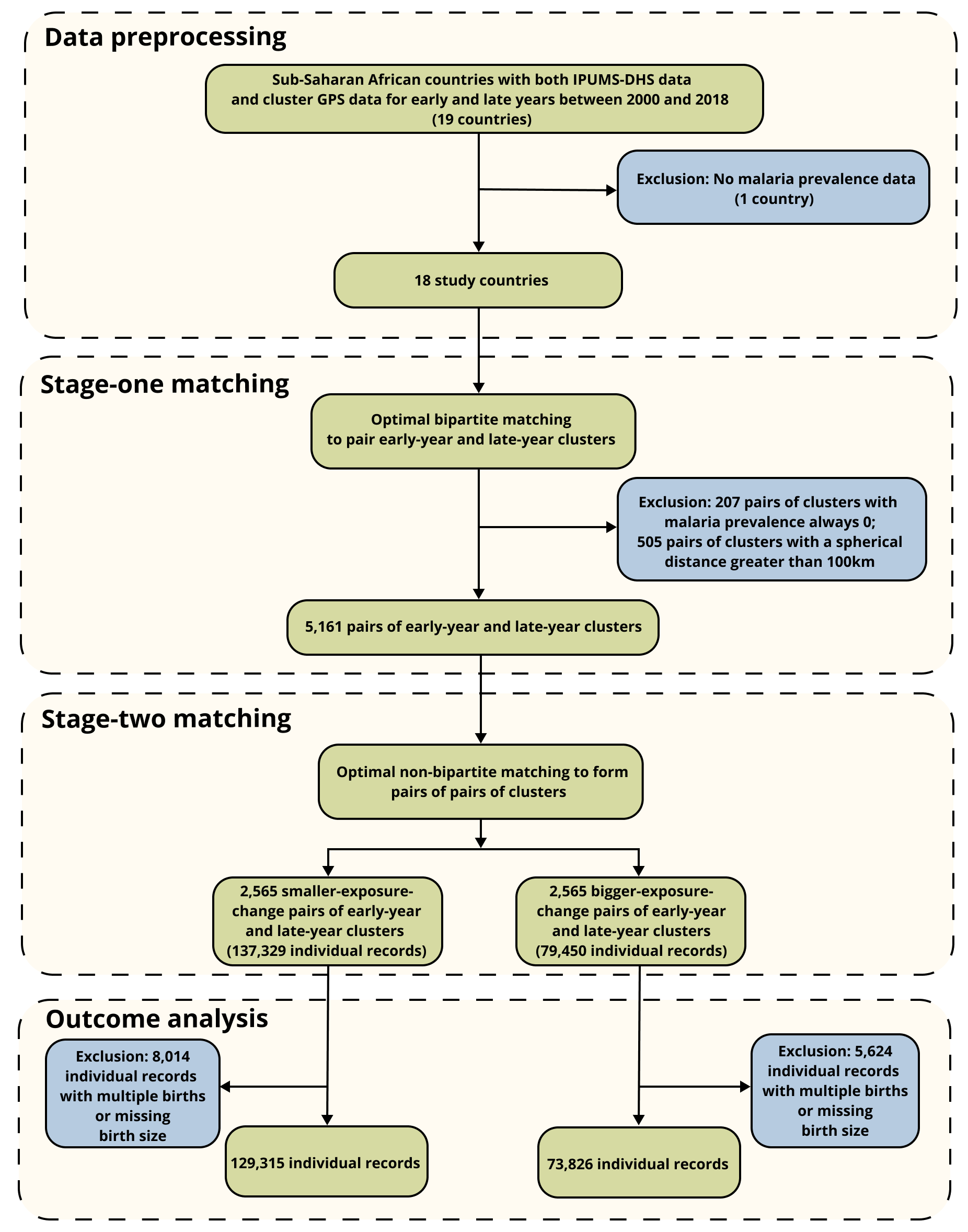}
    \label{fig: workflow}
\end{figure}

Under the post-matching working model (\ref{eqn: outcome model}), the effect coefficient $\beta_{1}$ being negative (or positive) means that, from early to late years, reduced malaria prevalence increases (or decreases) cluster-level average birthweight. We fit the post-matching model (\ref{eqn: outcome model}) among the 20 imputed datasets produced in Step 3 and use Rubin's rule \cite{rubin1996multiple} to combine the results across imputed datasets. The whole workflow of our statistical framework is also summarized in Figure~\ref{fig: workflow}.

\subsection{Step 5: Sensitivity Analysis for Unobserved Covariates}

To further explore the influences of potential unobserved covariates (those that could violate the parallel trend assumption even if the observed covariates have been adjusted for) on outcome analysis, we conduct sensitivity analysis for unobserved covariates under the omitted variable bias framework \cite{cinelli2020making}. Specifically, for the $j$-th pair of clusters in the $i$-th pair of pairs of clusters ($i=1,\dots, I$ and $j=1,2$), we let $u_{ij}^{\text{early}}$ (or $u_{ij}^{\text{late}}$) denote its unobserved covariate in the early-year (or the late-year). We then consider the following sensitivity analysis model that includes an additional term of the change in unobserved covariates over the early and late years, denoted as $u_{ij}^{\text{diff}}=u_{ij}^{\text{late}}-u_{ij}^{\text{early}}$, into the post-matching working model (\ref{eqn: outcome model}):
\begin{equation}\label{eqn: sensitivity model}
    Y_{ij}^{\text{diff}}= \beta_{0} +\beta_{1}Z_{ij}^{\text{diff}} + \gamma_{\text{early}}^{T}\mathbf{X}_{ij}^{\text{early}} + \gamma_{\text{late}}^{T}\mathbf{X}_{ij}^{\text{late}} + \sum_{i^{\prime}=1}^{I}c_{i^{\prime}} \mathbbm{1}\{i=i^{\prime}\} +\lambda u_{ij}^{\text{diff}} + \epsilon_{ij}.
\end{equation}
Under model (\ref{eqn: sensitivity model}), we calculate two measures for the sensitivity of our outcome analysis to unobserved covariates: the robustness value (RV) for the direction of effect (i.e., the sign of the effect coefficient $\beta_{1}$), denoted as $RV_{\text{sign}}$, and the robustness value for statistical significance (under level $\alpha=0.05)$ of the effect coefficient $\beta_{1}$, denoted as $RV_{\alpha = 0.05}$ \cite{cinelli2020making}. These two measures help clarify how strong an unobserved covariate needs to be to alter our quantitative conclusions. Specifically, the $RV_{\text{sign}}$ and $RV_{\alpha = 0.05}$ denote the degree of association (in terms of the proportion of residual variance explained) an unobserved covariate would need to have with the exposure and the outcome in order to (i) alter the direction of the exposure effect (i.e., the sign of the effect coefficient $\beta_{1}$) and (ii) render our effect coefficient statistically non-significant at the 0.05 significance level, respectively. Therefore, an $RV_{\text{sign}}$ close to 1 means that the direction of the exposure effect drawn from our analysis is robust enough to withstand a strong unobserved covariate. Conversely, an $RV_{\text{sign}}$ near 0 indicates that even a weak unobserved covariate could potentially overturn the direction of the exposure effect \cite{cinelli2020making}. Similarly, an $RV_{\alpha = 0.05}$ close to 1 (or near 0) means that the statistical significance detected from our outcome analysis is robust to even a strong unobserved covariate (or sensitive to even a weak unobserved covariate).

\section{Results}\label{sec: results}

\subsection{Stage-One Matching (Step 1)}\label{subsec: results of stage-one matching}

We assess the geographical closeness of paired early-year and late-year clusters by examining (i) their geographic proximity, measured by the mean spherical distance, elevation difference, within-pair longitude and latitude correlations, and the average longitude and latitude; (ii) similarity in $\text{\textit{Pf}PR}_{2-10}$, in both early and late years, between the paired DHS clusters. We summarize all these quantities in Table~\ref{tab: examine step 1}, which shows that our stage-one matching yields geographically proximate cluster pairs with similar malaria prevalence trajectories across early and late years. Specifically, among the 5,161 pairs of early-year and late-year clusters, the within-pair longitude and latitude correlations are nearly perfect (i.e., near 1). The average spherical distance between the paired early-year and late-year clusters is 14.235 km, with an average elevation difference of 1.452 meters. The average values of $\text{\textit{Pf}PR}_{2-10}$, in both early and late years, are very similar between the paired clusters. In Figure \ref{fig: The result of the bipartite matching, taking Kenya as an example}, we use Kenya as an example to visually demonstrate geographic closeness between paired early-year clusters (represented by red points) and late-year clusters (represented by blue points), further demonstrating the effectiveness of our stage-one matching.

\subsection{Stage-Two Matching (Step 2)}\label{subsec: stage-two matching}

Recall that stage-two matching yields 2{,}565 pairs of pairs of clusters. In each pair of pairs of clusters, between early and late years, one pair of clusters experiences a sharper reduction in $\text{\textit{Pf}PR}_{2-10}$ (referred to as ``bigger-exposure-change pair of clusters") and one pair of clusters experiences a less evident change in $\text{\textit{Pf}PR}_{2-10}$ (denoted as ``smaller-exposure-change pair of clusters"). We report the covariate balance (i.e., sociodemographic and socioeconomics similarity) between the paired bigger-exposure-change and smaller-exposure-change pairs of clusters in Table~\ref{tab:covariate balance}, which shows the mean values of covariates for bigger-exposure-change and smaller-exposure-change pairs after matching, as well as the standardized covariate differences after matching (i.e., the difference in mean covariate values between the paired bigger-exposure-change and smaller-exposure-change pairs of clusters standardized by the standard deviation of covariate values before matching~\cite{rosenbaum2010design}). Table~\ref{tab:covariate balance} shows that, in both early and late years, all the 12 covariates are well balanced after stage-two matching, with the absolute standardized differences of all the covariates less than 0.1 ~\cite{rosenbaum2010design, zubizarreta2012using, zhang2023social}. In addition to covariate balance, Table~\ref{tab:covariate balance} also reports the post-matching mean value of changes in $\text{\textit{Pf}PR}_{2-10}$ between the early and late years among the bigger-exposure-change pairs and that among the smaller-exposure-change pairs, with the absolute standardized difference equaling 0.953. This indicates sufficient heterogeneity in changes in malaria prevalence for detecting potential exposure effects from a difference-in-differences study.

\subsection{Missing Birthweights Imputation (Step 3)}

In Table S2 in the supplementary materials, we report the summary of the imputation model based on linear Bayesian regression. We can see that in our imputation model, mother's reported birth size, mother's age, child's birth order, and mother's education are important predictors, which agrees with previous literature \cite{fraser1995association, richards2001birth, de2004risk}. According to the model diagnosis for our Bayesian imputation model, the R-hat statistics for all predictors are approximately 1.00, indicating satisfactory model convergence. Furthermore, both the bulk effective sample size (ESS) and the tail ESS exceed 400, suggesting sufficient sampling efficiency of the Bayesian imputation model \cite{smith1973general, mitchell1988bayesian}.

\subsection{Outcome Analysis (Step 4)}

The results of our outcome analysis are summarized in Table~\ref{tab: outcome analysis1}. The estimated effect coefficient (i.e., the $\beta_{1}$ in model (\ref{eqn: outcome model})) of the difference in $\text{\textit{Pf}PR}_{2-10}$ (i.e., the $Z_{ij}^{\text{diff}}$) on the difference in average birthweight in grams (i.e., the $Y_{ij}^{\text{diff}}$) is $-155.747$ (95\% CI: $[-250.073, -61.420]$; $p$-value $<$ 0.001). This evidence supports our hypothesis that reduction in malaria prevalence can increase infants’
birthweights over the study period 2000–2018 at the community level. For example, among the pairs of clusters we study, the largest reduction in $\text{\textit{Pf}PR}_{2-10}$ over early and late years is 0.635, which is estimated to increase the average birthweight by $-155.747 \times -0.635 \approx 98.899$ grams (95\% CI: $[39.002, 158.796]$). Previous studies have found that an increase of 100 grams in birthweight is associated with reduced risks of various adverse birth and life-course outcomes such as neonatal mortality~\cite{watkins2016all}, life-course all-cause mortality~\cite{risnes2011birthweight}, low muscle strength~\cite{dodds2012birth}, low bone mineral density~\cite{baird2011does}, and leukaemia~\cite{caughey2009birth}. Meanwhile, according to Table~\ref{tab: outcome analysis1}, our outcome analysis also detects evident associations between changes in birthweights and variations in the mother's education, methods of contraception, mother's age, child's birth order, household's wealth index, child's sex, and mother's marital status, which is consistent with existing literature~\cite{murphy1984employment, teitelman1990effect, messer2010invited, jefferis2002birth, blencowe2019national}.

\subsection{Sensitivity Analysis (Step 5)}

Over the 20 imputed datasets, the average $RV_{\text{sign}}=9.4\%$, which means that to alter the sign (i.e., direction) of the point estimate of the effect coefficient $\beta_{1}$ in our outcome analysis, an unobserved covariate (orthogonal to the observed covariates) needs to explain more than 9.4\% of the residual variance of both the exposure and the outcome. Also, the average $RV_{\alpha = 0.05}=3.8\%$, indicating that to overturn the statistical significance of the effect coefficient $\beta_{1}$ in our outcome analysis, an unobserved covariate (orthogonal to the observed covariates) needs to explain more than 3.8\% of the residual variance of both the exposure and the outcome. For reference, the residual variance of outcome explained by the difference in mother's education is $0.6\%$, that explained by the difference of child’s sex is $1.1\%$, and that explained by the difference in mother’s marital status is $ 0.7\%$. Since both $RV_{\text{sign}}$ and $RV_{\alpha = 0.05}$ are higher than any of the quantities provided as references above, confounding factors as strong as the covariates mentioned above cannot fully eliminate the point estimate.



\section{Discussion}\label{sec: discussion}

In this work, we propose a novel statistical approach (i.e., a pair-of-pairs design using two-stage matching based on optimal bipartite and non-bipartite matching) for conducting a difference-in-differences study with survey time and location heterogeneity (e.g., DHS data) and continuous exposure (e.g., malaria prevalence). By leveraging the heterogeneity of changes in malaria prevalence across sub-Saharan African regions, we apply it to evaluate the community-level impact of reduced malaria prevalence on infants' birthweights in sub-Saharan Africa. We find that each percentage point of reduction in malaria transmission rate (measured by $\text{\textit{Pf}PR}_{2-10}$) is estimated to increase community-level average birthweight by $155.747\times 0.01 \approx 1.557$ grams (95\% CI: $[0.614, 2.501]$; $p$-value $<$ 0.001). This indicates that if a holoendemic region \cite{kevin2002epidemiological} (defined as $\text{\textit{Pf}PR}_{2-10}>0.75$) transitions to a hypoendemic region \cite{kevin2002epidemiological} (defined as $\text{\textit{Pf}PR}_{2-10}<0.1$), the average community-level birthweight would increase by approximately $155.747\times (0.75-0.1) \approx 101.236$ grams. Existing research suggests that an increase of 100 grams in birthweight is statistically significantly associated with improvements in various birth and life-course outcomes, such as reduction in neonatal mortality risk~\cite{watkins2016all} and life-course all-cause mortality risk~\cite{risnes2011birthweight}, increment in muscle strength~\cite{dodds2012birth} and bone mineral density in lumbar spine~\cite{baird2011does}, and reduced risk of leukaemia~\cite{caughey2009birth}. To our knowledge, this is the first study that detects strong evidence of the community-level effect of reduced malaria prevalence on improving the outcome of birthweights in sub-Saharan Africa under a quasi-experimental design.

Strengths of our study include the development and application of a novel quasi-experimental design -- a pair of pairs design based on a two-stage matching procedure using bipartite and non-bipartite matching -- on a large and representative dataset that includes 203{,}141 individual records from 18 sub-Saharan countries over the study period 2000 -- 2018. Using optimal bipartite matching in our stage-one matching, we have successfully addressed time and location heterogeneity in the DHS datasets, laying the groundwork for a difference-in-differences study. Furthermore, by employing optimal non-bipartite matching, we have enhanced matching quality without the need to dichotomize a continuous exposure (e.g., $\text{\textit{Pf}PR}_{2-10}$), thus preserving more observations, significantly improving covariate balance after matching, and introducing the flexibility of allowing the effect size to scale with degrees of change in exposure dose (e.g., $\text{\textit{Pf}PR}_{2-10}$). After the two-stage matching procedure, our outcome analysis and sensitivity analysis have detected strong evidence about the impact of changes in malaria burden on birthweight at the community level.

Our research has both scientific and methodological implications. First, although some previous studies \cite{mayor2015changing} have found reductions in infants' birthweights among malaria-infected women with declined antimalarial immunity (e.g., may be observed in communities with reduced malaria prevalence), our study is the first one that provides strong evidence that overall reduction in exposure to malaria infection during pregnancy outweighs these individual-level changes in birthweights among malaria-infected pregnancies. Given that birthweight is a commonly used proxy for overall birth outcomes, our study provides robust evidence to motivate future studies and investments in mitigating the global malaria burden. Second, our research provides a general statistical framework -- a pair of pairs design based on two-stage bipartite and non-bipartite matching -- for conducting a difference-in-differences study when facing the challenges of survey time and location heterogeneity and continuous exposures, so it has strong potential to benefit many other epidemiological studies.

This study is not without limitations. First, for each sub-Saharan African country included in the study, we utilize data from only two years (early and late years). This limitation arises because the DHS survey years are not continuous. While the advantage of this design is that it allows us to maximize the time gap between the study periods, enabling us to observe sufficient heterogeneity in changes in malaria prevalence across different regions, we cannot account for the actual intervention years between the early and late years. Second, although we have included 12 household-level and individual-level sociodemographic and socioeconomic covariates in our matching and outcome analysis, unobserved covariates may still exist. To address this limitation, we have conducted a sensitivity analysis for unobserved covariates, and the outcome analysis results remain robust.

\section*{Acknowledgement}

The authors thank Dylan Small for his helpful comments and suggestions. 

\bibliographystyle{unsrt}

\bibliography{Epidemiology}

\clearpage

\begin{figure}
\caption{Illustration of the difference between bipartite matching and non-bipartite matching. }
\begin{center}
\includegraphics[width=0.7\linewidth]{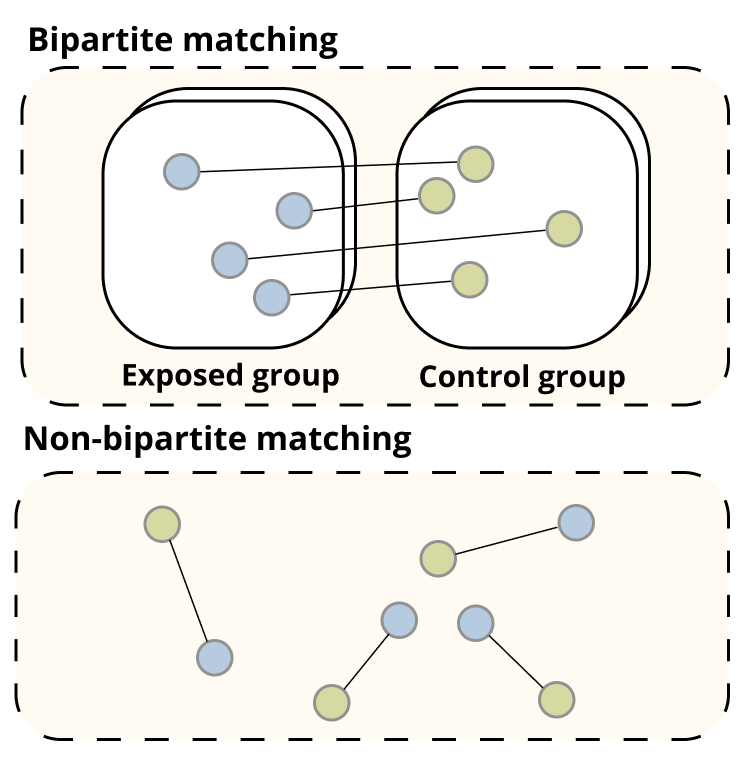}
\end{center}
\label{fig: Difference between bipartite matching and non-bipartite matching}
\end{figure}

\begin{figure}
\caption{The result of stage-one matching based on optimal bipartite matching, taking Kenya as an example.}
\begin{center}
\includegraphics[width=1 \linewidth]{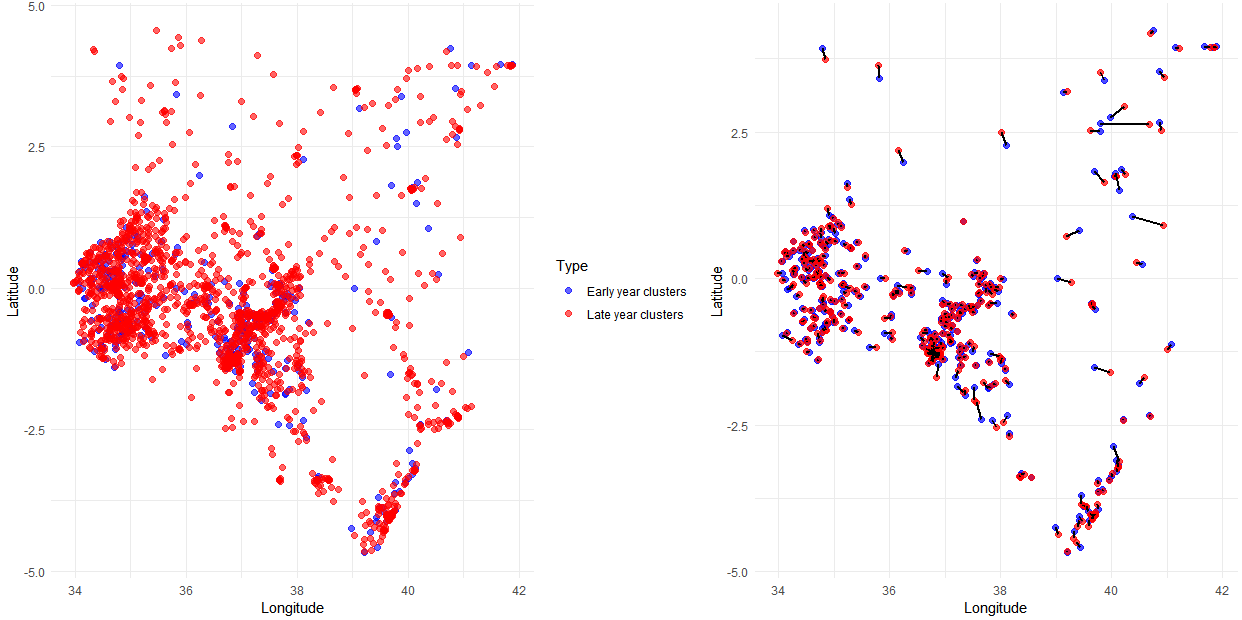}
\end{center}
\label{fig: The result of the bipartite matching, taking Kenya as an example}
\end{figure}

\clearpage

\begin{table}[ht]
\centering
  \caption{The 18 selected sub-Saharan African countries along with their chosen early and late years of malaria prevalence (i.e., annual $\text{\textit{Pf}PR}_{2-10}$ from the MAP data) and IPUMS-DHS early and late years. Note that some DHS datasets span over two successive years.}
 \begin{tabular}{l r r r r} 
 \hline
 & \multicolumn{2}{c}{Malaria prevalence}& \multicolumn{2}{c}{IPUMS-DHS} \\
  Country & Early-year & Late-year & Early-year & Late-year \\ [0.5ex] 
 \hline
 Benin & 2001 & 2012 & 2001 & 2012 \\ 
 Burkina Faso & 2003 & 2010-11 & 2003 & 2010 \\ 
 Burundi & 2010 & 2016-17 & 2010 & 2016 \\ 
 Cameroon  & 2004 & 2018-19 & 2004 & 2018  \\
 Congo Democratic Republic  & 2007 & 2013-14 & 2007 & 2013  \\
 Cote d’Ivoire & 2000 & 2011 & 1999 & 2011  \\
 Ghana & 2003 & 2014 & 2003 & 2014 \\
 Guinea  & 2005 & 2018 & 2005 & 2018 \\ 
 Kenya  & 2003-04 & 2014 & 2003 & 2014 \\
 Liberia  & 2007 & 2013 & 2007 & 2013  \\
 Mali   & 2001 & 2018 & 2001 & 2018 \\
 Namibia  & 2000 & 2013 & 2000 & 2013 \\
 Nigeria  & 2003 & 2018 & 2003 & 2018 \\
 Rwanda & 2004 & 2014-15 & 2005 & 2014 \\
 Senegal  & 2005 & 2017 & 2005 & 2017 \\
 Tanzania  & 2000 & 2015-16 & 1999 & 2015 \\
 Uganda  & 2001 & 2016 & 2001 & 2016 \\
 Zambia  & 2007 & 2018 & 2007 & 2018 \\
 \hline   
\end{tabular}
\label{tab: years}
\end{table}

\begin{table}[ht]
  \caption{Sociodemographic and socioeconomic covariates considered in our stage-two matching based on optimal non-bipartite matching. We include the covariates for both early-year and late-year.}
\small
\centering
 \begin{tabular}{l r} 
 \hline
 \textbf{Covariate} & \textbf{Categories} \\ [0.5ex] 
 \hline
 Household electricity & 0 = dwelling has no electricity; 1 = otherwise \\ 
 Household main material of floor & 1 = natural or earth-based; 2 = rudimentary; 3 = finished \\ 
 Household toilet facility & 0 = no facility; 1 = with toilet \\ 
 Living in urban or rural area & 0 = rural; 1 = urban \\ 
 Mother's education & 0 = no education; 1 = primary; 2 = secondary or higher \\ 
 Indicator of use of modern contraception & 0 = no; 1 = yes \\ 
 Mother's age & Numeric \\ 
 Child's birth order & 1 = first child; 2 = second, third, or fourth child; 3 $\geq$ fifth child \\ 
 Household wealth index & 1 = poorest; 2 = poorer; 3 = middle; 4 = richer; 5 = richest \\ 
 Child's sex & 0 = female; 1 = male \\ 
 Mother's current marital or union status & 0 = never married; 1 = married or living together \\ 
 Antenatal care indicator & 0 = no; 1 = yes \\ 
 \hline   
\end{tabular}
\label{tab:variables_step2}
\end{table}

\begin{table}[ht]

\caption{Descriptive statistics of the results of stage-one matching.}
\centering 

  \resizebox{\textwidth}{!}{%
  \begin{tabular}{lrrrr}
  \hline
    & \multicolumn{2}{c}{Correlation} & \multicolumn{2}{c}{Mean}\\
    \hline
   & Longitude &	Latitude & Spherical distance & Elevation difference
 \\
    Clusters within pairs &	1.000	& 1.000	& 14.235km	& 1.452m
 \\
   \hline
   & \multicolumn{2}{c}{Mean} & \multicolumn{2}{c}{Mean}\\
   \hline
   &  Longitude &  Latitude  &  $\text{\textit{Pf}PR}_{2-10}$ (early) &  $\text{\textit{Pf}PR}_{2-10}$ (late)\\ 
    Early-year clusters &	10.442	& 3.755	&0.309	&0.230 \\
    Late-year clusters &	10.454 &	3.756 &	0.329	& 0.226 \\
    \hline
 \end{tabular}
 }
 \label{tab: examine step 1}
\end{table}

\begin{table}[ht]

\caption{Covariate balance after non-bipartite matching (stage-two matching). For both the bigger-exposure-change and smaller-exposure-change pairs, we report the mean value of each covariate (in both early and late years) after non-bipartite matching. SEC pairs donates smaller-exposure-change pairs, and BEC pairs donates bigger-exposure-change pairs. We also report the standardized differences (Std.diff) of each covariate between the bigger-exposure-change and smaller-exposure-change pairs.}

\setlength{\tabcolsep}{2pt}
\centering
{%
  \begin{tabular}{lrrrrr}
  \hline
  & SEC pairs & BEC pairs & Std.diff \\ 
   & (2{,}565 pairs) & (2{,}565 pairs) &  \\
  \hline
Change in exposure ($\text{\textit{Pf}PR}_{2-10}^{\text{late}}-\text{\textit{Pf}PR}_{2-10}^{\text{early}}$) & 0.002 & -0.166 & 0.953 \\
\hline
Covariates  &   &   &   \\
Urban/rural (early) & 0.376 & 0.372 & 0.007 \\
Urban/rural (late) & 0.403 & 0.408 & -0.010 \\
Toilet facility (early) & 0.269 & 0.262 & 0.019 \\
Toilet facility (late) & 0.397 & 0.404 & -0.017 \\
Floor material (early) & 1.900 & 1.890 & 0.016 \\
Floor material (late) & 2.080 & 2.090 & -0.016 \\
Electricity (early) & 0.261 & 0.253 & 0.022 \\
Electricity (late) & 0.395 & 0.402 & -0.018 \\
Mother's education (early) & 0.687 & 0.542 & -0.030 \\
Mother's education (late) & 0.805 & 0.845 & -0.071 \\
Modern method of contraception (early) & 0.705 & 0.723 & -0.031 \\
Modern method of contraception (late) & 0.838 & 0.882 & -0.071 \\
Mother's age (early) & 29.040 & 29.100 & -0.011 \\
Mother's age (late) & 29.300 & 29.300 & 0.027 \\
Child's birth order (early) & 2.080 & 2.080 & -0.014 \\
Child's birth order (late) & 2.020 & 2.020 & 0.028 \\
Household wealth index (early) & 3.100 & 3.080 & 0.014 \\
Household wealth index (late) & 3.050 & 3.100 & -0.035 \\
Child's sex (early) & 0.508 & 0.508 & -0.001 \\
Child's sex (late) & 0.510 & 0.510 & 0.001 \\
Mother's current marital or union status (early) & 0.884 & 0.884 & -0.002 \\
Mother's current marital or union status (late) & 0.858 & 0.856 & 0.014 \\
Antenatal care (early) & 0.633 & 0.627 & 0.035 \\
Antenatal care (late) & 0.624 & 0.620 & 0.015 \\

  \hline
  \end{tabular}
} 
\label{tab:covariate balance}
\end{table}

\begin{table}[ht]
\centering
\caption{Results of the outcome analysis based on multiple imputation.}
\begin{tabular}{lrrr}
  \hline
Regressor & Est. & 95\% CI & \textit{p}-value \\ 
  \hline
Change in exposure ($\text{\textit{Pf}PR}_{2-10}^{\text{late}}-\text{\textit{Pf}PR}_{2-10}^{\text{early}}$)  & -155.747 & [-250.073, -61.420] & $<$0.001\\ 
Urban/rural (early) & 43.739 & [-19.216, 106.693] & 0.173  \\ 
Urban/rural (late) & 5.224 & [-56.051,  66.499] & 0.867\\ 
Toilet facility (early) & -43.816 & [-305.925, 218.293] & 0.743\\ 
Toilet facility (late) & 208.824 & [-114.782, 532.431] & 0.206\\ 
Floor material (early) & -23.862 & [-71.473, 23.749] & 0.326 \\ 
Floor material (late) & -35.408 & [-84.099, 13.283] & 0.154\\ 
Electricity (early) & 122.966 & [-151.169, 397.102] & 0.379\\ 
Electricity (late) & -171.809 & [-503.707, 160.089]  & 0.310 \\
Mother’s education (early) & -174.369 & [-557.475, 208.738] & 0.372 \\
Mother’s education (late) & 359.949  & [52.583, 667.315] & 0.022 \\
Modern method of contraception (early) &  125.665 & [-240.192, 491.523]  &  0.501 \\
Modern method of contraception (late) & -366.197 & [-655.429, -76.965] & 0.013 \\
Mother’s age (early) & -0.104 & [-10.125, 9.917] & 0.984 \\  
Mother’s age (late) & 9.815 & [-0.720, 20.350] & 0.068 \\  
Child’s birth order (early) & -128.329 & [-238.212, -18.446] & 0.022 \\   
Child’s birth order (late) &  -120.477 & [-232.630, -8.324] & 0.035 \\  
Household wealth index (early) & -30.163 & [-65.503, 5.176] & 0.094 \\  
Household wealth index (late) & 2.171  & [-32.360, 36.702] & 0.902 \\  
Child’s sex (early) & -218.913  & [-347.959, -89.868] & $<$0.001 \\  
Child’s sex (late) & 91.897 & [-55.880, 239.674] & 0.223\\  
Mother’s current marital or union status (early) & 143.001 & [-1.955, 287.957]  &  0.053 \\ 
Mother’s current marital or union status (late)  & -158.923 & [-302.006, -15.840]  & 0.029 \\ 
Antenatal care (early) & -24.612 & [-148.209, 98.986] & 0.696 \\ 
Antenatal care (late) & -57.966  & [-139.228, 23.297] & 0.162 \\ 
   \hline
\end{tabular}
\label{tab: outcome analysis1}
\end{table}

\clearpage

\section*{Appendix A: Technical Details of Optimal Bipartite and Non-bipartite Matching}

\subsection*{A.1: Optimal Bipartite Matching (Stage-One Matching)}

Consider country $A$ with $m$ survey clusters in the early year. Each cluster is represented by a vector of geographic coordinates $(x_{i}^{\text{early}}, y_{i}^{\text{early}})$, where $x_{i}^{\text{early}} $ denotes the longitude and $ y_{i}^{\text{early}}$ denotes the latitude, for $i = 1, 2, 3, \dots, m$. Meanwhile, suppose that country $A$ has $n$ survey clusters in the late year, where the geographic coordinates are represented as $(x_{j}^{\text{late}}, y_{j}^{\text{late}})$, in which $x_{j}^{\text{late}}$ is the longitude and $y_{j}^{\text{late}}$ is the latitude, for $j = 1, 2, 3, \dots, n$. Let $\mathbf{r}_{x}^{\text{\text{early}}}=(r_{x, 1}^{\text{\text{early}}}, \dots, r_{x, m}^{\text{\text{early}}})$ denote the vector of rank-transformed longitudes of clusters of country $A$ in the early year, where $r_{x, i}^{\text{early}}=\sum_{i^{\prime}=1}^{m}\mathbbm{1}\{x_{i}^{\text{early}}\geq x_{i^{\prime} }^{\text{early}}\}$ for $i=1,\dots, m$. Let $\mathbf{r}_{y}^{\text{\text{early}}}=(r_{y, 1}^{\text{\text{early}}}, \dots, r_{y, m}^{\text{\text{early}}})$ denote the vector of rank-transformed latitudes of clusters of country $A$ in the early year, where $r_{y, i}^{\text{early}}=\sum_{i^{\prime}=1}^{m}\mathbbm{1}\{y_{i}^{\text{early}}\geq y_{i^{\prime} }^{\text{early}}\}$ for $i=1,\dots, m$. Let $\mathbf{r}_{x}^{\text{\text{late}}}=(r_{x, 1}^{\text{\text{late}}}, \dots, r_{x, n}^{\text{\text{late}}})$ denote the vector of rank-transformed longitudes of clusters of country $A$ in the late year, where $r_{x, j}^{\text{late}}=\sum_{j^{\prime}=1}^{n}\mathbbm{1}\{x_{j}^{\text{late}}\geq x_{j^{\prime} }^{\text{late}}\}$ for $j=1,\dots, n$. Let $\mathbf{r}_{y}^{\text{\text{late}}}=(r_{y, 1}^{\text{\text{late}}}, \dots, r_{y, n}^{\text{\text{late}}})$ denote the vector of rank-transformed latitudes of clusters of country $A$ in the late year, where $r_{y, j}^{\text{late}}=\sum_{j^{\prime}=1}^{n}\mathbbm{1}\{y_{j}^{\text{late}}\geq y_{j^{\prime} }^{\text{late}}\}$ for $j=1,\dots, n$. Consider the sample covariance matrix
\begin{equation*}
   \widehat{\Sigma}=\begin{pmatrix}
\widehat{\sigma}^{2}_{11} & \widehat{\sigma}^{2}_{12} \\
\widehat{\sigma}^{2}_{21} & \widehat{\sigma}^{2}_{22}
\end{pmatrix},
\end{equation*}
where $\widehat{\sigma}^{2}_{11}$ denotes the sample variance of $(\mathbf{r}_{x}^{\text{\text{early}}}, \mathbf{r}_{x}^{\text{\text{late}}})$, the $\widehat{\sigma}^{2}_{12}=\widehat{\sigma}^{2}_{21}$ denotes the sample covariance between $(\mathbf{r}_{x}^{\text{\text{early}}}, \mathbf{r}_{x}^{\text{\text{late}}})$ and $(\mathbf{r}_{y}^{\text{\text{early}}}, \mathbf{r}_{y}^{\text{\text{late}}})$, and $\widehat{\sigma}^{2}_{22}$ denotes the sample variance of $(\mathbf{r}_{y}^{\text{\text{early}}}, \mathbf{r}_{y}^{\text{\text{late}}})$. Then, the rank-based Mahalanobis distance~\cite{mclachlan1999mahalanobis} $D(i,j)$ between clusters $(x_{i}^{\text{early}}, y_{i}^{\text{early}})$ and $(x_{j}^{\text{late}}, y_{j}^{\text{late}})$ is defined as:
\begin{equation*}
D(i,j) = \sqrt{(r_{x, i}^{\text{\text{early}}}-r_{x, j}^{\text{\text{late}}}, r_{y, i}^{\text{\text{early}}}-r_{y, j}^{\text{\text{late}}}) \Sigma^{-1} (r_{x, i}^{\text{\text{early}}}-r_{x, j}^{\text{\text{late}}}, r_{y, i}^{\text{\text{early}}}-r_{y, j}^{\text{\text{late}}})^\top}.
\end{equation*}
The benefits of using rank-based Mahalanobis distance to measure distances are: (i) Using rank-transformed values reduces the impact of particularly high or low longitude or latitude values (i.e., outliers) on the overall calculation~\cite{gonzalez2016new}; (ii) Mahalanobis distance reflects not only the individual variations of longitude and latitude but also their covariances (joint variations) and adjusts the distance calculation based on them~\cite{xiang2008learning}.

In the context of our two-stage matching procedure, the objective of the optimal bipartite matching is to find a specific pair-matching combination within each country, such that each of the $m$ early-year clusters is paired with one of the $n$ late-year clusters if $m\leq n$, or each of the $n$ late-year clusters is paired with one of the $m$ early-year clusters if $m>n$~\cite{rosenbaum1989optimal, hansen2006optimal}, with the minimal sum of the rank-based Mahalanobis distances between the paired early-year and late-year clusters, subject to the constraint that each distance is smaller than some prespecified caliper $ c $~\cite{rosenbaum1989optimal, hansen2006optimal}. Without loss of generality, we assume $m\leq n$ (otherwise, we just need to switch the index of the early and late-year clusters). Let $\mathcal{J}$ be an injective map from $\{1, \dots, m\}$ to $\{1, \dots, n\}$. Then, our stage-one matching based on optimal bipartite matching can be represented as the following optimization problem:
\begin{equation*}
    \underset{\mathcal{J}}{\text{minimize}} \quad \sum_{i=1}^{m}\big[ D(i, \mathcal{J}(i)) + \rho\times \mathbbm{1}\{D(i,\mathcal{J}(i)) \geq c\} \big],
\end{equation*}
where $\rho$ is some prespecified large number. We apply the aforementioned optimal bipartite matching to each of the 18 selected Sub-Saharan African countries, using \texttt{optmatch} package ~\cite{hansen2007optmatch} in \texttt{R}. The caliper $c$ is set to 0.2 times the standard deviation of the absolute differences in rank-based Mahalanobis distances between early-year and late-year clusters.

\subsection*{A.2: Optimal Non-Bipartite Matching (Stage-Two Matching)}

In the optimal bipartite matching (stage-one matching), we obtained 5{,}161 pairs of early-year and late-year clusters across all the 18 study countries. In our optimal non-bipartite matching procedure (stage-two matching), the objective is to pair pairs of early-year and late-year clusters such that the selected 12 covariates (in both early and late years) of these pair of pairs are as similar as possible while their difference in changes in $\text{\textit{Pf}PR}_{2-10}$ (between the early and late years) is as large as possible. Let us consider the $s$-th pair of early-year and late-year clusters, denoted as $p_s$, where $s = 1, 2, \dots, 5{,}161$. The change in exposure of $p_s$ is $Z_s^{\text{diff}} = (\text{\textit{Pf}PR}_{2-10}^{\text{late}} - \text{\textit{Pf}PR}_{2-10}^{\text{early}})_s$, which represents the change in $\text{\textit{Pf}PR}_{2-10}$ between the early and late years. Let $\mathbf{X}_s=(\mathbf{X}_{s}^{\text{early}}, \mathbf{X}_{s}^{\text{late}})$ represent the 12 selected covariates (in both early and late years) of the $s$-th pair of early-year and late-year clusters. Because optimal non-bipartite matching \cite{lu2001matching, lu2011optimal} aims to generate pairings with balanced observed covariates as well as large differences in patterns of exposure dose (in our case, the differences in changes in exposure dose $Z_s^{\text{diff}}$), the distance between the $s$-th pair of early-year and late-year clusters, $p_s$, and the $s^{\prime}$-th pair of early-year and late-year clusters, $p_{s^{\prime}}$, is calculated using the following formula:
\begin{equation*}
 \widetilde{D}(s,s^{\prime}) = d_{M}(\mathbf{X}_s, \mathbf{X}_{s^{\prime}})  + \rho^{\prime}\times \mathbbm{1}\{d_{M}(Z_{s}^{\text{diff}}, Z_{s^{\prime}}^{\text{diff}})<\xi\},
\end{equation*}
in which $ d_{M}(\mathbf{X}_s, \mathbf{X}_{s^{\prime}})$ denotes the Mahalanobis distance between $\mathbf{X}_s$ and $\mathbf{X}_{s^{\prime}}$. The penalty term $\rho^{\prime}\times \mathbbm{1}\{d_{M}(Z_s^{\text{diff}}, Z_{s^{\prime}}^{\text{diff}})<\xi\}$ represents that we will add a penalty for proximity in changes in exposure between the pairs of clusters $s$ and $s^{\prime}$, employed to prevent the matching of two pairs of clusters with similar change in exposure across time. In this study, $\rho^{\prime}$ is set to 1000 and $\xi$ is set to 0.05. 

For implementation, we use the \texttt{gendistance} function and the \texttt{distancematrix} function from \texttt{R}'s \texttt{nbpMatching} package~\cite{lu2011optimal} to calculate the pairwise distances $\widetilde{D}(s,s^{\prime})$ for all 5{,}161 pairs of clusters. We then perform pairwise matching on these 5{,}161 pairs of clusters, with the objective of minimizing the sum of distances between each matched pair of pairs of clusters. This matching algorithm is based on the Derig algorithm~\cite{derigs1988solving}, and we can implement this matching process using the \texttt{nonbimatch} function from \texttt{R} package \texttt{nbpMatching} ~\cite{lu2011optimal}.

\section*{Appendix B: More Details on Missing Data Imputation}

We utilize the \texttt{brm} function in $\texttt{R}$ package \texttt{brms}~\cite{burkner2017brms} to fit the Bayesian linear regression~\cite{mitchell1988bayesian} and use it for multiple imputation. Bayesian models can capture the uncertainty of model parameters by introducing prior distributions and the likelihood of the data~\cite{zhou2010note}. \texttt{brms} constructs a Bayesian linear regression model and uses all non-missing data to estimate the posterior distribution of the model parameters~\cite{zhou2010note, burkner2017brms}. For missing birthweight data, \texttt{brms} employs Markov Chain Monte Carlo (MCMC) to sample these missing values from the posterior distribution~\cite{hastings1970monte}, which is based on the observed data and the model parameters. This process accounts for the correlation between covariates and the inherent uncertainty in the data~\cite{hastings1970monte, burkner2017brms}. For each MCMC chain, different possible imputations for the missing values are drawn~\cite{burkner2017brms}. Repeating this process generates multiple sets of imputed results, thereby reflecting the uncertainty in the missing data.

We select the following covariates in Table~\ref{tab:variable_step3} to fit the Bayesian linear regression: the size of the infant at birth as reported subjectively by the mother (1 = very small or smaller than average; 2 = average; 3 = larger than average or very large), the mother's age in years, the infant's birth order number (1 = the first infant born to a mother; 2 = the second, third, or fourth infant born to a mother; 3 = otherwise), the household wealth index (1 = poorest; 2 = poorer; 3 = middle; 4 = richer; 5 = richest), whether the mother is living in an urban or rural area (0 = rural; 1 = urban), the mother's educational level (0 = no education; 1 = primary; 2 = secondary or higher), the infant's sex (0 = female; 1 = male), the mother's current marital or union status (0 = never married or formerly in union; 1 = married or living together), and an indicator of whether the mother receives any antenatal care while the infant is in utero (0 = no; 1 = yes). Because the effects of a mother's age and the infant's birth order on birthweight may not be linear \citep{selvin1971four}, we also add quadratic terms for both the mother's age in years and the infant's birth order to the regression.

We fit a Bayesian linear regression model over the 203,141 individual records with the default weakly informative prior in the \texttt{brm} function ~\cite{gelman2008weakly}. These weakly informative priors have been shown to outperform Gaussian and Laplacian priors in various scenarios~\cite{gelman2008weakly}. Specifically, the regression intercept and coefficient terms are assigned a Student's t-distribution with 3 degrees of freedom, a location parameter of 3200, and a scale parameter of 593. The residual standard deviation is also modeled with a Student's t-distribution with 3 degrees of freedom, centered at 0, and with a scale parameter of 593. The posteriors of the regression are estimated using the MCMC method, with two Markov chains and a total of 1,000 iterations per chain, including the warm-up phase~\cite{burkner2017brms}. Table~\ref{tab: bayes linear regress1} presents the results of the fitted Bayesian linear regression model (i.e., the imputation model). The Gelman-Rubin diagnostic values (Rhat) are all close to 1, indicating good model convergence. The effective sample size indicators (Bulk ESS and Tail ESS) are all much greater than 400. According to the fitted imputation model, the significant predictors for missing birthweight include the mother’s age, the infant’s birth order, urban or rural, the mother’s educational level, the infant's sex, the mother's marital status, indicator of antenatal care, and the mother’s reported infant's birth size.

\section*{Appendix C: More Details on Sensitivity Analysis}

We use the \texttt{sensemakr} package~\cite{cinelli2020making} in \texttt{R} to conduct a sensitivity analysis for unobserved covariates that could violate the conditional parallel trends assumption. For each of the 20 imputed datasets, we apply the model presented in Step 5 in the main text for sensitivity analysis and calculate the corresponding $RV_{\text{sign}}$ and $RV_{\alpha = 0.05}$, of which the results are presented in Table~\ref{tab: Sensitivity analysis to unobserved covariates}. Although we cannot directly assess the strength of an unobserved covariate, we can still make relative assessments about its strength by comparing it with the strength of observed covariates~\cite{cinelli2020making}. To facilitate such comparison, we introduce an additional assumption that the early-year and late-year covariates contribute equally to the outcome. We select the change in cluster-level mother's educational level, the change in cluster-level rates of infant’s sex, and the change in cluster summaries of mother’s marital status as the reference observed covariates. We find that, among the datasets after two-stage matching, the residual variance of the change in outcome explained by the change in cluster-level mother's educational level is 0.6\%, those explained by the change in cluster-level rates of infant’s sex is 1.1\%, and that explained by the change in cluster summaries of mother’s marital status is 0.7\%. Combining these with the sensitivity analysis results shown in Table~\ref{tab: Sensitivity analysis to unobserved covariates}, our calibrated sensitivity analysis indicates that the evidence detected in our outcome analysis is robust to potential unobserved covariates.

\section*{Appendix D: A Survey of the Mechanisms of Malaria Prevalence on Birthweight}

\subsection*{D.1: Individual-Level Malaria Infection and Birthweight}

During pregnancy, \textit{Plasmodium falciparum} tends to sequester in the placenta~\cite{teo2016functional, tran2020impact}. The sequestration of infected red blood cells by the malaria parasite and immune cell infiltration result in placental thickening, which alters nutrient and waste exchange between the mother and fetus, thereby leading to reduced birthweight~\cite{chua2021malaria, bakken2021impact} (see the top half of Figure~\ref{fig: appendixplt}). Specifically, (i) malaria infection is associated with dysregulation of placental vascular growth factors and metabolic hormones, impairing trophoblast differentiation, which may result in increased umbilical artery resistance and poor placental perfusion~\cite{abrahams2005role, elphinstone2019early}; (ii) women with \textit{Plasmodium falciparum} positive episodes during pregnancy exhibit an increased ratio of angiopoietin-1 (Ang-1) to angiopoietin-2 (Ang-2), and persistent Ang-2 levels during pregnancy may lead to the formation of new vessels while inhibiting the maturation of existing vessels. This is in contrast to healthy pregnancies, where Ang-1 and Ang-2 levels gradually increase and decrease, respectively, throughout pregnancy, promoting initial branching and subsequent maturation of the vasculature, thus inhibiting fetal growth and development~\cite{geva2002human, silver2010dysregulation}; (iii) the placentas of women infected with malaria during early pregnancy show signs of irreversible damage at delivery, including reduced transport villi, increased syncytial knotting, and placental lesions~\cite{crocker2004syncytiotrophoblast, chaikitgosiyakul2014morphometric}, which also limits the fetus's ability to obtain nutrients, thereby reducing birthweight.

\subsection*{D.2: Reduced Community-Level Malaria Prevalence and Reduced Antimalarial Immunity}

After exposure to malaria, pregnant women can typically generate two types of antibodies: (i) IgG antibodies against the CS2 parasite, specifically targeting the \textit{Plasmodium falciparum} strain CS2~\cite{muanda2015antimalarial, cutts2020pregnancy}. The CS2 parasite expresses the pregnancy-specific VAR2CSA protein, which binds to chondroitin sulfate A in the placenta, allowing infected red blood cells to accumulate in the placenta and leading to placental malaria~\cite{keitany2022invariant}; (ii) IgG antibodies against non-pregnancy-specific antigens, which are typically produced after an individual is exposed to \textit{Plasmodium falciparum} or other Plasmodium species~\cite{gbaguidi2024igg}. When individuals are bitten by infected mosquitoes or come into contact with the parasite, the immune system recognizes these non-pregnancy-specific antigens and initiates the production of IgG antibodies to combat the infection~\cite{dharmaratne2022quantification}. Unlike antibodies targeting pregnancy-specific antigens such as VAR2CSA, non-pregnancy-specific IgG antibodies can be generated in all infected individuals, including men, non-pregnant women, and children~\cite{yimam2021systematic}.

\begin{figure}
\caption{Illustration of the mechanisms of the possible positive and negative impacts of reduced malaria prevalence on birthweight.}
\begin{center}
\includegraphics[width=1\linewidth]{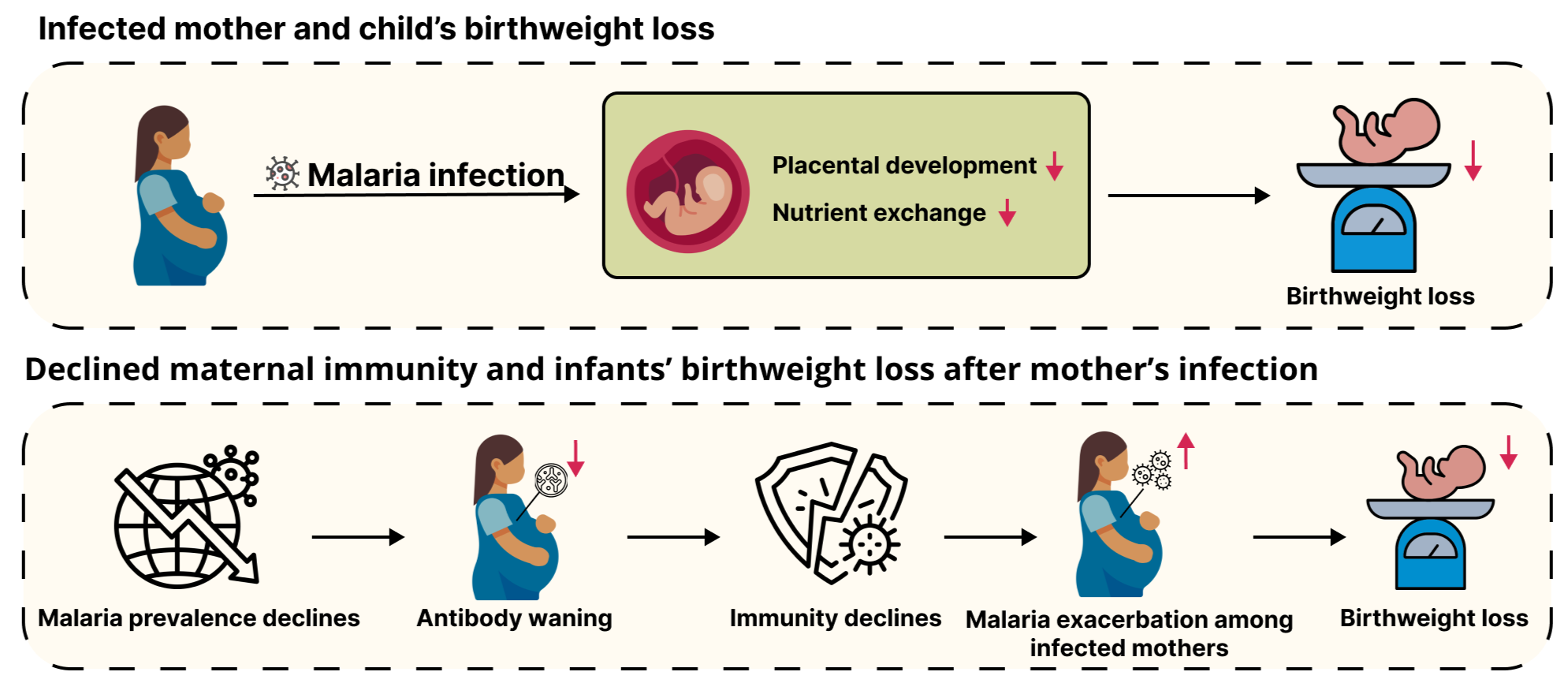}
\end{center}
\label{fig: appendixplt}
\end{figure}

Malaria can re-emerge in areas where transmission has ceased or is at very low levels, but immunity to malaria requires sustained exposure and stimulation to be maintained~\cite{cohen2012malaria, mayor2015changing}. Current research shows that within five years of declining malaria prevalence, the levels of IgG antibodies against CS2 parasites and non-pregnancy-specific IgG antibodies can decrease by more than twofold due to reduced exposure in pregnant women~\cite{mayor2015changing}. The decline in immunity increases parasite density in cases of infection~\cite{mayor2015changing}, exacerbating the severity of infection and having a more pronounced negative impact on infants' birthweight~\cite{heng2021relationship}. We illustrate the process in the bottom half of Figure~\ref{fig: appendixplt}.


\clearpage

\clearpage
\begin{table}[ht]
  \caption{Sociodemographic covariates for multiple imputation. We include the covariates for both early year and late year. For mother’s age and infant’s birth order, we also include their corresponding quadratic terms in the imputation model.}
\small
\centering
 \begin{tabular}{l r} 
 \hline
 \textbf{Covariate} & \textbf{Categories} \\ [0.5ex] 
 \hline 
  Living in urban or rural area & 0 = rural; 1 = urban \\ 
 Mother's education & 0 = no education; 1 = primary; 2 = secondary or higher \\ 
 Mother's age & Numeric \\ 
 Infant's birth order number & 1 = first infant; 2 = second, third, or fourth infant; 3 = otherwise \\ 
 Household wealth index & 1 = poorest; 2 = poorer; 3 = middle; 4 = richer; 5 = richest \\ 
 Infant's sex & 0 = female; 1 = male \\ 
 Infant's birth size & 0 = smaller than average; 1 = bigger than or equal to average \\ 
 Mother's current marital or union status & 0 = never married; 1 = married or living together \\ 
 Antenatal care indicator & 0 = no; 1 = yes \\ 
 \hline   
\end{tabular}
\label{tab:variable_step3}
\end{table}

\begin{table}[ht]
\centering
\caption{Summary of the Bayesian linear regression model (i.e., the imputation model) fitted among individual records with observed birthweight.}
\small
\setlength{\tabcolsep}{2pt}
\begin{tabular}{lrrrrrr}
  \hline
Predictor & Est. & S.E. & 95\% CI & Rhat & Bulk ESS & Tail ESS \\ 
  \hline
(Intercept) & 2853.41 & 34.48 & [2785.79, 2918.47] & 1.00 & 689 & 685 \\ 
Mother's age (linear term) & 8.10 & 2.44 & [3.49, 12.97] & 1.00 & 556 & 522 \\ 
Mother's age (quadratic term) & -0.17 & 0.04 & [-0.25, -0.09] & 1.01 & 553 & 491 \\ 
Wealth index & 0.79 & 1.73 & [-2.62, 4.06] & 1.00 & 770 & 515 \\ 
Infant's birth order (linear term) & 52.33 & 3.41 & [45.76, 59.42] & 1.00 & 910 & 518 \\ 
Infant's birth order (quadratic term) & -2.69 & 0.33 & [-3.34, -2.06] & 1.00 & 844 & 482 \\ 
0 - rural; 1 - urban & -20.19 & 4.44 & [-29.04, -11.64] & 1.00 & 726 & 700 \\ 
Mother's education & 77.52 & 2.73 & [72.28, 82.63] & 1.01 & 1081 & 684 \\ 
Infant is boy & 60.58 & 3.51 & [53.68, 67.58] & 1.01 & 1163 & 537 \\ 
Mother is married or living together & -24.09  & 5.65 & [-34.81, -13.34] & 1.00 & 674 & 609 \\
Indicator of antenatal care & -73.57  & 4.11  & [-81.62, -65.45] & 1.00 & 1260 & 606 \\
Indicator of low birth size & -571.80 & 5.51 & [-582.12, -560.75] & 1.00 & 1023 & 565 \\ 
Indicator of large birth size & 572.40 & 4.16 & [564.20, 580.28] & 1.01 & 1190 & 616 \\
   \hline
\end{tabular}
\label{tab: bayes linear regress1}
\end{table}

\begin{table}[htbp]
\small
\centering
  \caption{Sensitivity analysis results for unobserved covariates among 20 imputed datasets.}
  \label{tab: Sensitivity analysis to unobserved covariates}
\begin{tabular}{lccccc}
\hline
Change in Exposure  & Est. & S.E. & $t$-value & $RV_{\text{sign}}$ & $RV_{\alpha = 0.05}$  \\ 
\hline 
$\text{\textit{Pf}PR}_{2-10}^{\text{late}}-\text{\textit{Pf}PR}_{2-10}^{\text{early}}$ & -155.847 & 48.111 & -3.239 & 9.4\% & 3.8\% \\ 
$\text{\textit{Pf}PR}_{2-10}^{\text{late}}-\text{\textit{Pf}PR}_{2-10}^{\text{early}}$  & -155.852 & 48.101 & -3.240 & 9.4\% & 3.8\%\\ 
$\text{\textit{Pf}PR}_{2-10}^{\text{late}}-\text{\textit{Pf}PR}_{2-10}^{\text{early}}$  & -155.494 & 48.079 & -3.234 & 9.4\% & 3.8\% \\ 
$\text{\textit{Pf}PR}_{2-10}^{\text{late}}-\text{\textit{Pf}PR}_{2-10}^{\text{early}}$  & -156.947 & 48.162 & -3.259 & 9.5\% & 3.9\% \\ 
$\text{\textit{Pf}PR}_{2-10}^{\text{late}}-\text{\textit{Pf}PR}_{2-10}^{\text{early}}$  & -155.670 & 48.102 & -3.236 & 9.4\% & 3.8\% \\
$\text{\textit{Pf}PR}_{2-10}^{\text{late}}-\text{\textit{Pf}PR}_{2-10}^{\text{early}}$  & -156.305 & 48.125 & -3.248 & 9.4\% & 3.8\% \\
$\text{\textit{Pf}PR}_{2-10}^{\text{late}}-\text{\textit{Pf}PR}_{2-10}^{\text{early}}$  & -155.749 & 48.125 & -3.238 & 9.4\% & 3.8\% \\
$\text{\textit{Pf}PR}_{2-10}^{\text{late}}-\text{\textit{Pf}PR}_{2-10}^{\text{early}}$  & -156.321 & 48.150 & -3.247 & 9.4\% & 3.8\% \\ 
$\text{\textit{Pf}PR}_{2-10}^{\text{late}}-\text{\textit{Pf}PR}_{2-10}^{\text{early}}$  & -154.310 & 48.112 & -3.207 & 9.3\% & 3.7\% \\ 
$\text{\textit{Pf}PR}_{2-10}^{\text{late}}-\text{\textit{Pf}PR}_{2-10}^{\text{early}}$  & -156.320 & 48.141 & -3.247 & 9.4\% & 3.8\%\\ 
$\text{\textit{Pf}PR}_{2-10}^{\text{late}}-\text{\textit{Pf}PR}_{2-10}^{\text{early}}$  & -156.125 & 48.121 & -3.244 & 9.4\% & 3.8\% \\ 
$\text{\textit{Pf}PR}_{2-10}^{\text{late}}-\text{\textit{Pf}PR}_{2-10}^{\text{early}}$  & -155.860 & 48.099 & -3.240 & 9.4\% & 3.8\% \\ 
$\text{\textit{Pf}PR}_{2-10}^{\text{late}}-\text{\textit{Pf}PR}_{2-10}^{\text{early}}$  & -154.750 & 48.134 & -3.215 & 9.3\% & 3.7\% \\ 
$\text{\textit{Pf}PR}_{2-10}^{\text{late}}-\text{\textit{Pf}PR}_{2-10}^{\text{early}}$  & -155.383 & 48.111 & -3.230 & 9.4\% & 3.8\% \\ 
$\text{\textit{Pf}PR}_{2-10}^{\text{late}}-\text{\textit{Pf}PR}_{2-10}^{\text{early}}$  & -156.093 & 48.131 & -3.243 & 9.4\% & 3.8\% \\ 
$\text{\textit{Pf}PR}_{2-10}^{\text{late}}-\text{\textit{Pf}PR}_{2-10}^{\text{early}}$  & -156.035 & 48.148 & -3.241 & 9.4\% & 3.8\% \\ 
$\text{\textit{Pf}PR}_{2-10}^{\text{late}}-\text{\textit{Pf}PR}_{2-10}^{\text{early}}$  & -154.657 & 48.145 & -3.212 & 9.3\% & 3.7\% \\
$\text{\textit{Pf}PR}_{2-10}^{\text{late}}-\text{\textit{Pf}PR}_{2-10}^{\text{early}}$  & -155.894 & 48.119 & -3.240 & 9.4\% & 3.8\% \\
$\text{\textit{Pf}PR}_{2-10}^{\text{late}}-\text{\textit{Pf}PR}_{2-10}^{\text{early}}$  & -155.345 & 48.124 & -3.228 & 9.4\% & 3.8\% \\
$\text{\textit{Pf}PR}_{2-10}^{\text{late}}-\text{\textit{Pf}PR}_{2-10}^{\text{early}}$  & -155.971 & 48.128 & -3.241 & 9.4\% & 3.8\% \\
\hline 
\textbf{Average} & \textbf{-155.897} & \textbf{48.123} & \textbf{-3.236} & \textbf{9.4\%} & \textbf{3.8\%} \\
\hline 
\end{tabular}


\end{table}



\end{document}